\documentclass[11pt]{CGC2}
\usepackage[english]{babel}
\usepackage{verbatim}
\usepackage{psfrag}
\usepackage{graphicx}
\usepackage{epsfig}
\usepackage{xypic}
\usepackage{tikz}

\newcommand{\Sym}{\mathrm{Sym}}

\newcommand{\GL}{\mathrm{GL}}

\newcommand{\N}{\text{$\mathbf{N}$}}

\usepackage{bm,url}
\usepackage[utf8x]{inputenc}

\begin{document}

\Logo{Preprint 2010\ \ $\qquad$ CGC latex}

\begin{frontmatter}

\title{Do AES encryptions act randomly?}

{\author{Anna Rimoldi}} {{\tt (rimoldi@science.unitn.it)}}\\
{{Department of Mathematics, Univ. of Trento, Italy.}}

{\author{Massimiliano Sala}} {{\tt (maxsalacodes@gmail.com)}}\\
{{Department of Mathematics, Univ. of Trento, Italy.}}

{\author{Enrico Bertolazzi}} {{\tt (enrico.bertolazzi@ing.unitn.it)}}\\
{{Department of Mechanical and Structural Engineering, Univ. of Trento, Italy.}}

\runauthor{A.~Rimoldi, M.~Sala, E.~Bertolazzi}

\begin{abstract}
The Advanced Encryption Standard (AES) is widely recognized as the
most important block cipher in common use nowadays.
This high assurance in AES is given by its resistance to ten years of
extensive cryptanalysis, that
has shown no weakness, not even any deviation from the statistical
behaviour expected from a random permutation. Only reduced versions of
the ciphers have been broken,
but they are not usually implemented.
In this paper we build a distinguishing attack on the AES, exploiting
the properties of a novel cipher embedding. With our attack we give
some statistical evidence
that the set of AES-$128$ encryptions acts on the message space in a way
significantly different than that of the set of random permutations
acting on the same space.
While we feel that more computational experiments by independent third
parties are needed in order to validate our statistical results, we
show that the non-random behaviour
is the same as we would predict using the property of our embedding.
Indeed, the embedding lowers the nonlinearity of the AES rounds and
therefore the AES encryptions tend, on average, to keep low the rank
of low-rank matrices constructed in the large space. 
Our attack needs $2^{23}$ plaintext-ciphertext pairs and costs the equivalent
of $2^{48}$ encryptions.

We expect our attack to work also for AES-$192$ and AES-$256$, 
as confirmed by preliminary experiments.

\end{abstract}
\end{frontmatter}

\section*{Introduction}
The Advanced Encryption Standard (AES) is widely recognized as the
most important block cipher in common use nowadays
\cite{CGC-misc-nistAESfips197}.
Its 256-bit version (AES256) can even be used at top secret level (\cite{CGC-misc-nsaAes}).
This high assurance in AES is given by its resistance to ten years of
extensive cryptanalysis: AES
has shown no weakness, not even any deviation from the statistical
behaviour expected from a random permutation.
Only reduced versions of the cipher have been broken,
but they are not usually implemented
(see e.g. \cite{CGC-cry-prep-RST10}, Section $2$).

For a high-security cipher it is essential that nobody can distinguish
its encryption functions from random functions.
It is not enough that the encryption function associated to a key
cannot be distinguished
from a random map (single-key attack), or that the encryption
functions associated to related
keys cannot be distinguished from a set of random maps.
A high security cipher must behave so randomly that it must be
impossible to distinguish (a random sample of)
the whole set of AES encryptions from (a random sample of) the set of
random permutations.

In this paper we build a special kind of distinguishing attack on the AES.
To be more precise, with our attack we give some statistical evidence
that the set of AES-$128$ encryptions acts on the message space in a way
different than that of the set of random permutations
acting on the same space.
In this paper we do not claim any other successful distinguishing
attack, neither single-key
nor related-keys.

Our attack has a subtle theoretical justification. We are able to
embed the AES (and actually also other translation-based
ciphers) in a larger cipher, as explained in full details in \cite{CGC-cry-prep-RST10}.
This embedding is designed to lower the
non-linearity of the AES rounds. The decrease in the non-linearity
should be noted by analysing the ranks
of some matrices (similarly to a Marsaglia Die-Hard test \cite{CGC-misc-NIST-SP80022}).
While we feel that more computational experiments by independent third
parties are needed in order to validate our statistical results, we
show that the non-random behaviour is the same as we would predict using the property of our embedding.
Indeed, we observe that the AES encryptions tend, on average, to keep
low the rank of low-rank matrices constructed in the large space.
This holds true apparently for all standard AES versions.

Our attack needs $2^{23}$ plaintext-ciphertext pairs and costs the equivalent
of $2^{48}$ encryptions, thanks to 
a highly specialized rank-computation algorithm.

The remainder of this paper is organized as follows.
In Section \ref{Preliminari} we sketch the internal structure of the AES, we
explain our embedding and we treat some statistical models related to statistical attacks.
In Section \ref{Attacco} we describe our attack strategy.
In Section \ref{Num_result} we report our attack numerical results, including
results on different AES versions.
 In Section  \ref{Computational} we discuss some
computational matters, presenting a rank-computation algorithm.
In Section \ref{Conclusioni} we provide our conclusions and several remarks.

\section{Preliminaries}
\label{Preliminari}
In this section we mainly follow the notations and the approach in \cite{CGC-cry-prep-RST10}, including viewing AES as a translation-based cipher.

\subsection{An AES description}
In this subsection we recall the essential structure of the AES cryptosystem viewed as translation-based (for a more traditional approach see \cite{CGC-cry-book-aesbook}).

Let $V=(\FF_2)^r$ with $r=128$  be the space of all possible messages (plaintexts or ciphertexts).
Let  $\mathcal{K}=(\FF_2)^{\ell}$  be the finite set of all possible keys (with $\ell=128, 192, 256$).
Any key $k \in \mathcal{K}$ specifies an encryption function $\phi_{k}$.
Let $x \in V$ be any plaintext. In order to obtain the corresponding ciphertext
$y=\phi_{k}(x)\in V$, the encryption proceeds through $N=10, 12, 14$ similar rounds, respectively (depending on $\ell$), as described below.

A preliminary translation via (addition with) the first round key $k^{(0)}$ in $(\FF_2)^{r}$ is applied to the plaintext to form the input to the first round ({\tt Round 1}). Other $N$ rounds follow.

Let $1\leq \rho \leq N-1$. 
A typical round ({\tt Round $\rho$}) can be written as the composition\footnote{Note that the order of the operation is exactly: $\gamma$, $\lambda$, and then $\sigma_k$.}
$\gamma \lambda \sigma_{k^{(\rho)}}$, where the map $\gamma$ is called {\sf SubBytes} and  works in parallel to each of the $16$ bytes of the data
({\sf SubBytes} is composed by two transformations: the inversion in $\FF_{2^8}$ and an affine transformation over $\FF_2$);
 the linear map $\lambda:V \rightarrow V$ is the composition of two linear operations known as  {\sf ShiftRows} and {\sf MixColumns};
 $\sigma_{k^{(\rho)}}$ is the translation with the round key $k^{(\rho)}$ (this operation is called {\sf AddRoundKey}).

The last round ({\tt Round $N$}) is atypical and can be written as $\gamma \bar{\lambda} \sigma_{k^{(N)}}$, where
the affine map $\bar{\lambda}$ is the {\sf ShiftRows} operation.

Obviously, we can see the linear operation $\lambda$ as a matrix ${\bf M}$. 
We observe that the order of $\lambda$ is quite small and equal to $8$:  $\lambda^8=1_V$.

\subsection{The embedding we are using}

We are interested in particular {\sf space embeddings} where the vector space $V=(\FF_2)^{r}$ and  $W$ is the vector space $(\FF_2)^{s}$, with $s > r$. 
We want to embed $V$ into $W$ by an injective map $\alpha$ and to extend $\phi_k \in \Sym(V)$ to a permutation
$\phi'_k \in \Sym(W)$ as shown in the following commutative diagram:
$$
 \xymatrix{\ar @{} [dr] |{\circlearrowright}
   V \ar[d]^{\phi_k} \ar[r]^{\alpha} & W \ar[d]^{\phi'_k}\\
   V \ar[r]^{\alpha} & W}
$$

In order to do this, we have to define the permutation $\phi'_k \in \Sym(W)$. We say that $\phi'_k$ is an {\sf extension} of $\phi_k$. 
Let $r=m b$, let $s=2^m b t$.
According to the setting described in \cite[Section 4]{CGC-cry-prep-RST10},
the space embedding $\alpha:V \rightarrow W$ we consider is defined as follows
\begin{equation}
\label{orbita}
\alpha(v)=(\varepsilon(v),\varepsilon ({\bf M} v),\dots, \varepsilon({\bf M}^{t-1} v))
\end{equation}
where:
\begin{enumerate}
 \item[a)] $\varepsilon :(\FF_{2^m})^b \rightarrow ((\FF_2)^{2^m})^{b}$ is a parallel map 
$\varepsilon(v_1,\dots,v_{b})=(\varepsilon '(v_1),\dots, \varepsilon '(v_{b}))$;
\item[b)] the map $\varepsilon ': \FF_{2^m} \rightarrow (\FF_2)^{2^m}$ is defined by means of a primitive element 
$\eta$ of $\FF_{2^m}$ as  
$$
\varepsilon '(0)=(1,\underbrace{0,\dots,0}_{2^m -1}) \qquad 
\varepsilon '(\eta^{i})=(0,\dots,0,\underset{\underset{i+1}{\uparrow}}{1},0,\dots,0) \quad \forall 1\leq i\leq 2^m-1 \,. 
$$ 
\item[c)] the matrix ${\bf M}$ in $\GL((\FF_{2})^{mb})$ has order $t$.
\end{enumerate}

Moreover, for  byte-oriented Mixing Layers, i.e. if ${\bf M} \in \GL((\FF_{2^m})^b)$, the following bound has been proved as Proposition 4.2  in  \cite{CGC-cry-prep-RST10}:
 $$\dim_{\FF_2}\big(\langle \mathrm{Im}(\alpha)\rangle \big)\leq 2^{m}bt -(bt-1)- mb(t-1) .$$

Let ${\bf M}:V \rightarrow V$ be the {\sf MixingLayer} of AES.
The map $\alpha:V \rightarrow W$ we propose for AES is defined 
as follows
\begin{equation}
\label{AESorbita}
\alpha(v)=(\varepsilon(v),\varepsilon ({\bf M} v),\dots, \varepsilon({\bf M}^7 v)),
\end{equation}
where:  $\varepsilon :(\FF_2)^{128}\rightarrow (\FF_2)^{4096}$, $\varepsilon ': \FF_{256} \rightarrow (\FF_2)^{256}$, ${\bf M}\in \GL((\FF_{2})^{128})$.\\
We have $t=8$, $b=16$ and $m=8$.
In Fact 4.4 (\cite{CGC-cry-prep-RST10}) we determined the dimension of $\langle\mathrm{Im}(\alpha)\rangle$, for $\alpha$ in (\ref{AESorbita}), 
using the fact that  ${\bf M} \in \GL((\FF_{256})^{16})$:

  $$\dim_{\FF_2}(\langle\mathrm{Im}(\alpha)\rangle)= 2^{m}bt -(bt-1)- mb(t-1)=31745.$$

The encryption $\phi_k$ is the composition of {\sf AddRoundKey}, {\sf Subbytes} and {\sf MixingLayer}.
So the only part of $\phi'_k$ which is not linear is the {\sf SubBytes} operation.\\

\begin{remark}
 The goal of our $\phi'_k$ construction is to have the non-linearity of {\sf SubBytes} decrease. 
\end{remark}

\subsection{On randomness and statistical distinguishers}

When a statistical test on data from a cryptographic algorithm is performed, we
wish to test whether the data “seem” random or not. It is impossible to design
a test that gives a decisive answer in all cases. However, there are many different properties of
randomness and non-randomness, and it is possible to design tests for these specific
properties.
An example of a set of tests is the NIST Test Suite \cite{CGC-misc-NIST-SP80022}. It is a statistical package consisting of $16$ tests that were developed to test the randomness of
(arbitrarily long) binary sequences produced by
cryptographic random or pseudorandom number generators. These tests focus on a
variety of different types of non-randomness that could exist in a sequence. 
For example, the  Marsaglia ``Die Hard'' test consists of determining whether the
statistics of ranks of $(32 \times 32)$  matrices over $\FF_2$, constructed with bits coming from the sequence,
agrees with the theoretical distributions.

We are going to introduce three cryptanalitic scenarios where such test can be applied. They are called ``distinguishing attacks'' or ``distinguishers''.

Generally speaking, distinguishing attacks against block ciphers aim at determining whether a given permutation corresponds to 
a random permutation\footnote{a permutation chosen uniformly at random from the set of all permutations.}
or to a $\phi_k$. 
Of course, there is always a distinguishing attack against any block cipher, since $|\mathcal{K}|< +\infty$, 
and so a brute-force key search will yield a distinguishing attack of average complexity $2^{\ell-1}$ (where $\ell$ is the key length),
but we are interested in attacks costing significantly less.

Let $m_1, \dots, m_N$ be some plaintexts, let $k$ be any key.
We denote by $\pi$ any random permutation in $\Sym((\FF_2)^r)$ and by $\phi_k$ the  encryption function for the key $k$; we have to consider the following situation, 
where one black box is involved and it contains\footnote{A weaker form of distinguisher assumes that the black box contain $\phi_k$ or $\pi$ with the same probability \cite{CGC-cry-art-Lucks96}.} either $\phi_k$ or $\pi$.

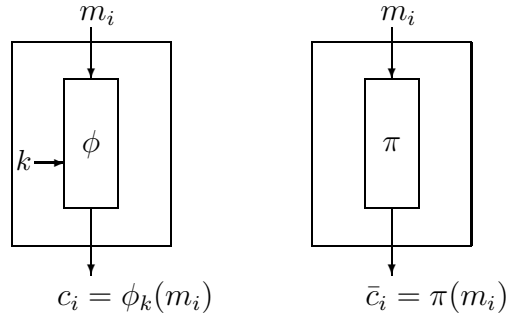
\begin{center}
\setlength{\unitlength}{1cm}
\begin{picture}(8,4.6)(0,0)
\put(0.3,1.5){\frame{\makebox(2.1,2.7){}}}
\put(1,2){\frame{\makebox(0.7,1.7){$\phi$}}}
\put(1.35,4.4){\vector(0,-1){0.68}}
\put(1.2,4.5){$m_i$}
\put(1.35,2){\vector(0,-1){0.9}}
\put(0.9,0.7){$c_i=\phi_k (m_i)$}
\put(0.35,2.5){\small{$k$}} 
\put(0.60, 2.6) {\vector(1,0){0.40}}
\put(4.3,1.5){\frame{\makebox(2.1,2.7){}}}
\put(5,2){\frame{\makebox(0.7,1.7){$\pi$}}}
\put(5.35,4.4){\vector(0,-1){0.68}}
\put(5.2,4.5){$m_i$}
\put(5.35,2){\vector(0,-1){0.9}}
\put(5,0.7){$\bar{c}_i=\pi (m_i)$}
\end{picture}
\end{center}

A {\sf single-key distinguishing attack} on a cipher $\mathcal{C}$ is any algorithm $\mathcal{A}$ able to distinguish the ciphertexts 
$\{c_{i}\}_{1\leq i\leq N}$ from the random ciphertexts $\{\bar{c}_{i}\}_{1\leq i\leq \N}$, using some information on the plaintexts.
There are two main variants: the chosen-plaintext and the known-plaintext. In both,
formally $\mathcal{A}$ takes as input a set of pairs $\{(m_1,\hat{c}_i),\ldots,(m_N,\hat{c}_N)\}$ where 
\begin{itemize} 
\item either   $\hat{c}_i=c_i\quad \forall 1 \leq i \leq N$,
\item or       $\hat{c}_i=\bar{c}_i\quad \forall 1 \leq i \leq N$,
\end{itemize}
and returns as output ``true'' or ``false'':
\begin{itemize}
\item $\mathcal{A}$ outputs ``true'' if and only if $\hat{c}_i=c_i$, $\forall i$ s.t. $1\leq i \leq N$.
\item $\mathcal{A}$ outputs ``false'' if and only if $\hat{c}_i=\bar{c}_i$, $\forall i$ s.t. $1\leq i \leq N$.
\end{itemize}

The difference between the two\footnote{There are other ways to consider the plaintexts, according to the possibilities 
and the capabilities of Eve.} variants:
\begin{itemize}
\item {\em Chosen-plaintext:} $\mathcal{A}$ can decide the plaintexts and obtain the corresponding ciphertexts.
In this case, such plaintexts are often chosen ``related'', i.e. satisfying some additional prescribed mathematical relations;
\item {\em Known-plaintext:} $\mathcal{A}$ cannot decide the plaintexts and we can only assume that $\mathcal{A}$ 
knows a certain amount of pairs (plaintext,ciphertext). In this case, the plaintexts are often supposed random.
\end{itemize}

If we have some \textbf{related} keys $k_1,\ldots,k_\tau$, we can describe a second cryptana\-li\-tic scenario. 
A {\sf related-key distinguishing attack} on a cipher $\mathcal{C}$ is any algorithm $\mathcal{A}$ 
able to distinguish the ciphertexts $\{c_{i,j}\}$ from the random ciphertexts $\{\bar{c}_{i,j}\}$, as in the following scheme.

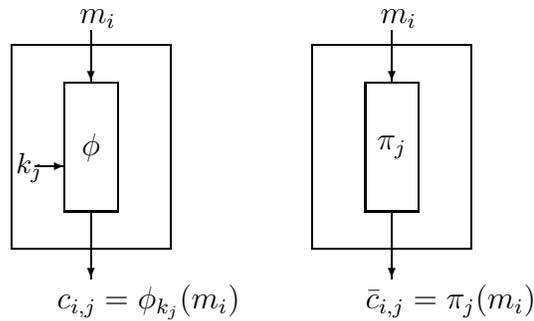
\begin{center}
\setlength{\unitlength}{1cm}
\begin{picture}(8,4.6)(0,0)
\put(0.3,1.5){\frame{\makebox(2.1,2.7){}}}
\put(1,2){\frame{\makebox(0.7,1.7){$\phi$}}}
\put(1.35,4.4){\vector(0,-1){0.68}}
\put(1.2,4.5){$m_i$}
\put(1.35,2){\vector(0,-1){0.9}}
\put(0.9,0.7){$c_{i,j}=\phi_{k_j} (m_i)$}
\put(0.35,2.5){\small{$k_j$}} 
\put(0.60, 2.6) {\vector(1,0){0.40}}
\put(4.3,1.5){\frame{\makebox(2.1,2.7){}}}
\put(5,2){\frame{\makebox(0.7,1.7){$\pi_j$}}}
\put(5.35,4.4){\vector(0,-1){0.68}}
\put(5.2,4.5){$m_i$}
\put(5.35,2){\vector(0,-1){0.9}}
\put(5,0.7){$\bar{c}_{i,j}=\pi_j (m_i)$}
\end{picture}
\end{center} 

\begin{remark}
In this model, $\mathcal{A}$ knows additionally some mathematical relations between
the keys used for encryption, but not the key values. 
Both the single key scenario and the related-key one describe a hypothetical situation, very difficult to reach in practice. 
Yet, a very secure block cipher must resist in both scenarios. \\
\end{remark}

There is another scenario where a distinguishing attack can be mounted.
This scenario is less common and we have not found an established name
in the literature for it, so we will call it
a {\sf random-key-sample distinguisher}.
As in the related-key scenario we consider some keys $k_1,\ldots,k_\tau$,
some plaintexts $\{m_1,\ldots,m_N\}$ and the corresponding
ciphertexts $\{c_{i,j}\}$. The difference is that now we consider the
keys as a random sample from the keyspace. This is a realistic assumption,
because when the session key is changed during transmissions a new
(pseudo)-random key is negiotiated between the two peers.
Formally, $\mathcal A$ behaves in the same way as the related-key algorithm, being able to
distinguish the set of actual encryptions $\{c_{i,j}\}$ from a set of random
vectors $\{\bar{c}_{i,j}\}$.
Clearly, also for the random-key-sample scenario we could have a
chosen-plaintext variant and a known-text variant, although it is rather
unlikely that a known-key attack can be succesful
(we would have both random keys and random plaintexts).

Our attack in the next section is of the third type. We use a strategy similar to that of  
the well-known Marsaglia test. 

\section{Strategy description}
\label{Attacco}
In this section, $\alpha$ is our embedding  (\ref{AESorbita}).\\
We recall that $\dim_{\FF_2}{(\langle \mathrm{Im}(\alpha)\rangle)}=31745$. Let $a^1, \ldots, a^{31745}$ be (not necessarily distinct) vectors in $V$.
Let $\mathcal{D}=\{a^1,a^2, \ldots, a^{31745}\}$, so  $|\mathcal{D}| \leq 31745$.
We construct the $(31745\times 2^{15})$-matrix $\mathbf{D}$ such that the $i$-th row is the image of the map 
$\alpha$ applied to the plaintext $a^i$, as in (\ref{matrice_orbita}).

\begin{eqnarray}
\label{matrice_orbita}
\mathbf{D}=
\left( \begin{array}{c}
\alpha(a^1) \\
\alpha(a^2) \\
\vdots \\
\alpha(a^{31745}) \\
\end {array} \right)
=
\left( \begin{array}{cccc}
\varepsilon(a^1) & \varepsilon({\bf M}a^1) & \cdots & \varepsilon({\bf M}^{7}a^1) \\
\varepsilon(a^2) & \varepsilon({\bf M}a^2) & \cdots & \varepsilon({\bf M}^{7}a^2) \\
\vdots & \vdots & \vdots & \vdots \\
\varepsilon(a^{31745}) & \varepsilon({\bf M}a^{31745}) & \cdots & \varepsilon({\bf M}^{7}a^{31745}) \\
\end {array} \right).
\end{eqnarray}

Let $\mathcal{M}$ be the set of all such matrices. Clearly, we have 
$$|\mathcal{M}|=(|V|)^{31745}=(2^{128})^{31745}.$$
We note that the weight of any row is $bt=128$.
\begin{center}
 What is the rank of $\mathbf{D}$ if $\mathcal{D}$ is random in $\mathrm{Im}(\alpha)$?
\end{center}

\noindent
The probability that a $\nu \times n$ random matrix $(\nu < n)$ with entries in $\FF_2$ has rank exactly
$s$ is significantly greater than the probability of having rank equal to $\nu-1$ or $\nu-2$ or less. 
On the other hand, for a square $n \times n$ random matrix in $\FF_2$ the rank $n-1$ is the most probable.
However, the most likely rank for $\mathbf{D}$ as in (\ref{matrice_orbita}) is not $31745$, although $31745< 2^{15}$,
because our construction imposes specific constraints, for example on the row weight. 
Let ${d_{\mathcal{M}}}$ denote the total number of matrices in $\mathcal{M}$ and let  $d_{31743}$ denote the number of all matrices in $\mathcal{M}$ 
with rank less than or equal to $31743$.
In \cite{CGC-cry-prep-RST10} we have computed the expected rank statistics for $\mathbf{D}$.
In particular, a direct consequence of Theorem~3.19 in \cite{CGC-cry-prep-RST10} is the following corollary:
\begin{corollary}
$$\frac{d_{31743}}{d_{\mathcal{M}}}= 0.1336357 \,.$$
\end{corollary}
Our attack is a \underline{random-key-sample distinguisher with chosen plaintext}, as detailed in the remainder of this section.\\

We choose $2^{16}$ plaintexts obtained by taking
$$
\bar{S}=\{ \bar{v}=(\bar{v}_1, \ldots, \bar{v}_{16}) \mid \bar{v} \in (\FF_{256})^{16}, \, \bar{v}_i=0, \, 1 \leq i\leq 14  \}.
$$
Clearly, $|\bar{S}|=(2^8)^2=2^{16}$.

\begin{remark}
We order $(\FF_2)^8$ following the order of the binary representation. For example, since $(00000010) \mapsto 2$ and  $(00001100)\mapsto 12$ we have
$(00001100)>(00000010)$.
We order $((\FF_2)^8)^2$ using the lexicographic ordering, induced by the previous order: $(a,b)>(a',b')$ if and only if either $a > a'$ or $a=a'$, $b>b'$.
Once chosen an irreducible polynomial $p \in \FF_2[x]$, with $\deg(p)=8$, we can identify $\FF_{256}$ with $(\FF_2)^8$ and so we can use the above orderings to order both $\FF_{256}$ and $(\FF_{256})^2$.
\end{remark}

Following the previous remark, we can write $\bar{S}=\{\bar{v}^1,\ldots, \bar{v}^{2^{16}}\}$ where $\bar{v}^{i+1}>\bar{v}^i$ for all $i$.
In other words, $\bar{S}$ is an ordered set of $2^{16}$ vectors.

We now describe an algorithm, that we call $\mathcal{B}$, that takes in input
an ordered set $S=\{v^1,\ldots,v^{2^{16}}\}$ of $2^{16}$ vectors in $V$
and that outputs a list of natural numbers $r_0,\ldots,r_{31745}$ computed
as follows.
We construct a first matrix $\mathbf{D}$ starting from $\{v^1,\ldots,v^{31745}\}$.
We compute the rank of  $\mathbf{D}$ and we store the value. 
We repeat this procedure with $\{v^2,\ldots, v^{31746}\}$ and so on with $\{v^{k+1},\ldots, v^{k+31745}\}$, where $2 < k \leq 33791$. 
In total, we compute the rank of $2^{16}-31745+1=33792$ matrices.
We define $r_j$ as the number of these matrices with rank $j$
(for $0\leq j\leq 31745$).
\\

We applied algorithm $\mathcal{B}$ to $\bar S$ and, since the rows of these matrices are strongly related (they share the same first $14$ bytes), 
we expect the corresponding ranks to be significantly lower than the most probable ones (see Subsection \ref{Fremarks} for details).

We can apply algorithm $\mathcal{B}$ to $\phi_k(\bar S)$ and to $\pi(\bar S)$,
where $\pi$ is any random map.
We would like to use the two output lists to distinguish between
$\phi_k$ and $\pi$, but we are not able to do it.
Instead, we choose a number $\tau$ and we do two different operations.
In one case, we apply $\mathcal{B}$ to $\phi_{k_i}(\bar S)$ for $\tau$ random
keys $k_1,\ldots,k_\tau$.
In the other case, we apply $\mathcal{B}$ to $\pi_i(\bar S)$ for $\tau$ random
maps $\pi_1,\ldots,\pi_\tau$.
In practice, we apply $\mathcal{B}$ to $\tau$ random ordered sets, each containing $2^{16}$ distinct vectors.\\
Finally, we use the output lists to distinguish between
$\{\phi_{k_i}\}$ and $\{\pi_i\}$.

We expect that the behaviour of the ranks coming from the encrypted matrices is distinguishable from the theoretical distribution, and in particular that these ranks are lower.
On the other hand, we expect that the ranks coming from the random matrices
follow the theoretical distribution.  
The results are reported in Section \ref{Num_result}.

\section{Numerical results}
\label{Num_result}

To mount the attack successfully we need to choose $\tau$ small enough
to allow for a practical computations {\em and} large enough to
overcome the effects induced by the variance in the random distribution.

Since we computed bunches of $10$ random
keys (and random maps), we observed that the values coming from the
random maps may be distinguishable
(from the expected distribution) if we consider up to $50$ maps. However,
with $70$ maps (or more) the random maps become undistinguishable,
especially compared to the drastic values obtained by the encryptions.

\pagebreak
Starting from a sample of $70$ matrices, we report the obtained rank values corresponding to 
 $r_0+\ldots+r_{31743}$ and $r_{31744}+r_{31745}$:

$$
\begin{tabular}{|c||c c c|}
\hline
\textrm{Rank}&\textrm{Random}& \textrm{AES}& \textrm{Expected}\\
\hline
\hline
     $ > 31743$ &  2049671 & 2047430 & 2049333 \\
 $\leq 31743$ &   315769 & 318010  &  316107    \\
\hline 
\end{tabular}
$$
\ \\

Now, we apply the $\chi ^2$ test between 
\begin{enumerate}
 \item Random and Expected, $\rightarrow$ P value equals ${\bf 0.51}$; 
 \item AES and Expected,  $\rightarrow$ P value equals ${\bf 0.0003}$.
\end{enumerate}

The lower the P-value, the higher the probability that the observed data do not come from the theoretical distribution.
It is customary in Statistics to consider $0.05$ as a threshold. Since the value for random data is $0.51$ and that for AES-$128$ is $0.0003$,
we may safely assume that, with high probability, the ranks observed for AES-$128$ do {\em not} come from a random sample.\\
Besides, apart from the threshold, the ratio between the two P values is remarkable. And the difference between the AES-$128$ ranks and
the theoretical distribution is exactly where we expect it to lie: in a significantly higher number of low-rank matrices.

In the following figure, we report the results of two samples coming from $\phi_k(\bar S)$ (the $70$ red dots) and
from $\pi(\bar S)$ (the $70$ blue circles). First, we ordered our samples according to the number of
low-rank matrices: on the left the samples with a smaller number
and on the right those with a larger number. Then we plotted
vertically this number. The horizontal line corresponds to the
expected value for low-rank matrices.
It should be apparent from the picture how the two groups of values are
separated.

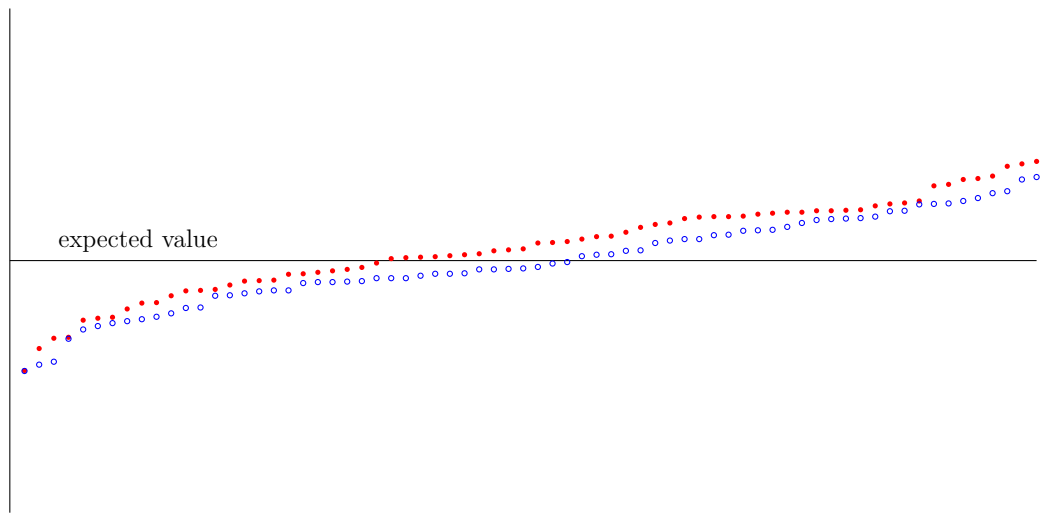
\begin{center}
\begin{tikzpicture}[scale=0.65]

\draw  (0,0) -- (0,10.32) ;
\draw  (0,5.16) -- (21,5.16) node [very near start, sloped,above,scale=0.8]{expected value};

\fill [red] (0.3,2.91) circle (1.5pt);
\fill [red] (0.6,3.36) circle (1.5pt);
\fill [red] (0.9,3.57) circle (1.5pt);
\fill [red] (1.2,3.59) circle (1.5pt);
\fill [red] (1.5,3.94) circle (1.5pt);
\fill [red] (1.8,3.98) circle (1.5pt);
\fill [red] (2.1,4.00) circle (1.5pt);
\fill [red] (2.4,4.17) circle (1.5pt);
\fill [red] (2.7,4.29) circle (1.5pt);
\fill [red] (3,4.30) circle (1.5pt);

\fill [red] (3.3,4.44) circle (1.5pt);
\fill [red] (3.6,4.54) circle (1.5pt);
\fill [red] (3.9,4.55) circle (1.5pt);
\fill [red] (4.2,4.57) circle (1.5pt);
\fill [red] (4.5,4.66) circle (1.5pt);
\fill [red] (4.8,4.74) circle (1.5pt);
      \fill [red] (5.1,4.75) circle (1.5pt);
\fill [red] (5.4,4.76) circle (1.5pt);
\fill [red] (5.7,4.88) circle (1.5pt);
\fill [red] (6,4.89) circle (1.5pt);

\fill [red] (6.3,4.92) circle (1.5pt);
      \fill [red] (6.6,4.95) circle (1.5pt);
\fill [red] (6.9,4.98) circle (1.5pt);
\fill [red] (7.2,5.02) circle (1.5pt);
\fill [red] (7.5,5.11) circle (1.5pt);
\fill [red] (7.8,5.20) circle (1.5pt);
\fill [red] (8.1,5.22) circle (1.5pt);
\fill [red] (8.4,5.23) circle (1.5pt);
\fill [red] (8.7,5.24) circle (1.5pt);
\fill [red] (9,5.26) circle (1.5pt);

\fill [red] (9.3,5.28) circle (1.5pt);
\fill [red] (9.6,5.30) circle (1.5pt);
\fill [red] (9.9,5.36) circle (1.5pt);
\fill [red] (10.2,5.38) circle (1.5pt);
\fill [red] (10.5,5.40) circle (1.5pt);
\fill [red] (10.8,5.52) circle (1.5pt);
\fill [red] (11.1,5.53) circle (1.5pt);
      \fill [red] (11.4,5.55) circle (1.5pt);
\fill [red] (11.7,5.60) circle (1.5pt);
      \fill [red] (12,5.65) circle (1.5pt);

\fill [red] (12.3,5.66) circle (1.5pt);
\fill [red] (12.6,5.74) circle (1.5pt);
\fill [red] (12.9,5.84) circle (1.5pt);
\fill [red] (13.2,5.90) circle (1.5pt);
\fill [red] (13.5,5.93) circle (1.5pt);
\fill [red] (13.8,6.02) circle (1.5pt);
      \fill [red] (14.1,6.05) circle (1.5pt);
\fill [red] (14.4,6.06) circle (1.5pt);
\fill [red] (14.7,6.06) circle (1.5pt);
\fill [red] (15,6.07) circle (1.5pt);

\fill [red] (15.3,6.11) circle (1.5pt);
\fill [red] (15.6,6.13) circle (1.5pt);
\fill [red] (15.9,6.15) circle (1.5pt);
\fill [red] (16.2,6.15) circle (1.5pt);
\fill [red] (16.5,6.18) circle (1.5pt);
\fill [red] (16.8,6.18) circle (1.5pt);
\fill [red] (17.1,6.19) circle (1.5pt);
\fill [red] (17.4,6.20) circle (1.5pt);
\fill [red] (17.7,6.28) circle (1.5pt);
\fill [red] (18,6.32) circle (1.5pt);

\fill [red] (18.3,6.34) circle (1.5pt);
\fill [red] (18.6,6.38) circle (1.5pt);
\fill [red] (18.9,6.69) circle (1.5pt);
\fill [red] (19.2,6.72) circle (1.5pt);
\fill [red] (19.5,6.82) circle (1.5pt);
\fill [red] (19.8,6.84) circle (1.5pt);
\fill [red] (20.1,6.89) circle (1.5pt);
\fill [red] (20.4,7.09) circle (1.5pt);
\fill [red] (20.7,7.14) circle (1.5pt);
\fill [red] (21,7.19) circle (1.5pt);


\draw[blue] (0.3,2.90) circle (1.5pt);
\draw[blue] (0.6,3.03) circle (1.5pt);
\draw[blue] (0.9,3.09) circle (1.5pt);
\draw[blue] (1.2,3.56) circle (1.5pt);
\draw[blue] (1.5,3.75) circle (1.5pt);
\draw[blue] (1.8,3.82) circle (1.5pt);
\draw[blue] (2.1,3.88) circle (1.5pt);
\draw[blue] (2.4,3.92) circle (1.5pt);
\draw[blue] (2.7,3.96) circle (1.5pt);
\draw[blue] (3,4.01) circle (1.5pt);
\draw[blue] (3.3,4.08) circle (1.5pt);
\draw[blue] (3.6,4.19) circle (1.5pt);
\draw[blue] (3.9,4.20) circle (1.5pt);
\draw[blue] (4.2,4.44) circle (1.5pt);
\draw[blue] (4.5,4.45) circle (1.5pt);
\draw[blue] (4.8,4.49) circle (1.5pt);
\draw[blue] (5.1,4.53) circle (1.5pt);
\draw[blue] (5.4,4.55) circle (1.5pt);
\draw[blue] (5.7,4.55) circle (1.5pt);
\draw[blue] (6,4.70) circle (1.5pt);
\draw[blue] (6.3,4.72) circle (1.5pt);

\draw[blue] (6.6,4.72) circle (1.5pt);
\draw[blue] (6.9,4.73) circle (1.5pt);
\draw[blue] (7.2,4.74) circle (1.5pt);
\draw[blue] (7.5,4.80) circle (1.5pt);
\draw[blue] (7.8,4.80) circle (1.5pt);
\draw[blue] (8.1,4.80) circle (1.5pt);
\draw[blue] (8.4,4.85) circle (1.5pt);
\draw[blue] (8.7,4.89) circle (1.5pt);
\draw[blue] (9,4.89) circle (1.5pt);
\draw[blue] (9.3,4.90) circle (1.5pt);
\draw[blue] (9.6,4.98) circle (1.5pt);
\draw[blue] (9.9,4.98) circle (1.5pt);
\draw[blue] (10.2,4.99) circle (1.5pt);

\draw[blue] (10.5,5.00) circle (1.5pt);
\draw[blue] (10.8,5.03) circle (1.5pt);
\draw[blue] (11.1,5.10) circle (1.5pt);
\draw[blue] (11.4,5.13) circle (1.5pt);
\draw[blue] (11.7,5.25) circle (1.5pt);
\draw[blue] (12,5.28) circle (1.5pt);
\draw[blue] (12.3,5.29) circle (1.5pt);
\draw[blue] (12.6,5.36) circle (1.5pt);
\draw[blue] (12.9,5.37) circle (1.5pt);
\draw[blue] (13.2,5.52) circle (1.5pt);
\draw[blue] (13.5,5.57) circle (1.5pt);
\draw[blue] (13.8,5.60) circle (1.5pt);
\draw[blue] (14.1,5.60) circle (1.5pt);
\draw[blue] (14.4,5.68) circle (1.5pt);
\draw[blue] (14.7,5.69) circle (1.5pt);
\draw[blue] (15,5.77) circle (1.5pt);
\draw[blue] (15.3,5.78) circle (1.5pt);
\draw[blue] (15.6,5.79) circle (1.5pt);
\draw[blue] (15.9,5.85) circle (1.5pt);
\draw[blue] (16.2,5.93) circle (1.5pt);
\draw[blue] (16.5,5.99) circle (1.5pt);

\draw[blue] (16.8,6.01) circle (1.5pt);
\draw[blue] (17.1,6.02) circle (1.5pt);
\draw[blue] (17.4,6.03) circle (1.5pt);
\draw[blue] (17.7,6.06) circle (1.5pt);
\draw[blue] (18,6.17) circle (1.5pt);
\draw[blue] (18.3,6.18) circle (1.5pt);
\draw[blue] (18.6,6.31) circle (1.5pt);
\draw[blue] (18.9,6.32) circle (1.5pt);
\draw[blue] (19.2,6.33) circle (1.5pt);
\draw[blue] (19.5,6.38) circle (1.5pt);
\draw[blue] (19.8,6.44) circle (1.5pt);
\draw[blue] (20.1,6.54) circle (1.5pt);
\draw[blue] (20.4,6.58) circle (1.5pt);
\draw[blue] (20.7,6.82) circle (1.5pt);
\draw[blue] (21,6.87) circle (1.5pt);
\end{tikzpicture}
\end{center}

\subsection{Furher remarks}
\label{Fremarks}
In this subsection we report on special applications of our algorithm
$\mathcal{B}$.
Some results are predictable:
\begin{itemize}
\item since our plaintext set $\bar{S}$ contains highly correlated vectors, we
     would expect that $\mathcal{B}$ outputs very low ranks when the input is
     $\bar{S}$ itself; indeed, in this case
     the output is $r_{4690}=33792$, that is, we get exactly $33792$
     ranks equal to $4690$. Actually, it is not difficult to prove that the dimension of the
     vector space generated by $\alpha(\bar{S})$ is $4821$, with arguments similar
     to those of the proof of Propostion 4.2 in \cite{CGC-cry-prep-RST10}. 
     Therefore, we would expect our $31745$ vector sample to form a matrix with a lower rank ($4690<4821$);

\item when we apply {\em one} round of AES (with any key) to $\bar S$, algorithm $\mathcal{B}$
     outputs again $r_{4690}=33792$. This may come as a surprise, but it is easily
     explained in our framework. One round\footnote{In the translation-based notation we are performing {\tt Round} $0$ and {\tt Round} $1$} means, in order, one key addition,
     one S-Box, one $\lambda$ and another key addition. Thanks to properties
     of the embedding $\alpha$, all the above operations are linear, except
     for the S-Box (see Proposition 4.5 in \cite{CGC-cry-prep-RST10}). However, the S-Box in this case
     does not change the type of subspace. Indeed, after the fist key-addition
     we have all vectors sharing the first $14$ coordinates and the last
     two are free to be any pair. Since the S-Box acts in parallel, it
     does not change this situation and so the dimension of the whole space and the ranks of our matrix remain unchanged;

\item things change when we apply {\em two} rounds of AES; the reason is that
     the MixColumns changes the structure, since it intermixes four bytes at a
     time; it is true that the MixColumns in the first round does not change
     the rank, but the change in the structure is fatal to the rank when
     the S-Box of the {\em second} round is applied; indeed, our experiments shows
     that $\mathcal{B}$ outputs in this case $r_{20548}=33792$. Again, this
     lower value is justified by the dimension of the $2$-round encryption of $\bar{S}$, which is $20679$;     

\item similarly, the rank raises with the application of {\em three} rounds:
     $\mathcal{B}$ outputs in this case $r_{31661}=33792$ and the dimension of the $3$-round encryption of $\bar{S}$ is $31681$;

\item when we apply {\em four} rounds or more, we get values near to the random
     setting (and the dimension of the subspace is $31745$, since it coincides with the whole space $\langle Im(\alpha)\rangle$); in this sense, we could say that the diffusion of AES
     is working from $4$ rounds and above, as it is usually agreed.
\end{itemize}

\pagebreak
\noindent
Some results are largely unexpected. For example:
\begin{itemize}
\item let us consider two random maps $\pi_1(\bar{S})$, $\pi_2(\bar{S})$ and two encryption functions $\phi_{k1}(\bar{S})$ and $\phi_{k2}(\bar{S})$ ($k_1 \not = k_2$).
Now, we apply algorithm $\mathcal{B}$ to the four corresponding  sets and we obtain the following rank distributions. 
$$
\begin{tabular}{|c||c c c c |c|}
\hline
\textrm{Rank}& $\textrm{Random}_1$ &$\textrm{Random}_2$ &$\textrm{AES}_1$& $\textrm{AES}_2$ & \textrm{Expected}\\
\hline
\hline
        $31745$ & 9782  &  9467& 9765 & 9554 & 9759 \\
     $ 31744$ & 19482  & 19765&  19525& 19569 & 19517\\
 $\leq 31743$ &  4528  & 4560&   4502&   4669 & 4516\\
\hline 
\end{tabular}
$$

As before, we apply the $\chi^2$ test between
\begin{enumerate}
 \item $\textrm{Random}_1$ and Expected, $\rightarrow$ P value equals $0.928$; 
 \item $\textrm{Random}_2$ and Expected,  $\rightarrow$ P value equals $0.002$;
 \item $\textrm{AES}_1$ and Expected,  $\rightarrow$ P value equals $0.97$;
 \item $\textrm{AES}_2$ and Expected,  $\rightarrow$ P value equals $0.008$.
\end{enumerate}

We note that case $(1)$ and case $(3)$ are statistically indistinguishable from the expected distribution, while case $(2)$ and $(4)$ appear statistically distinguishable from the expected.
In other words, if we apply our test {\em only to one key}\footnote{that is, if we try to mount a single-key attack.}, it fails badly, because it would distinguish with probability $0.5$, that is, it outputs a random value!
The reason for the single-key failure lies in the large variance among our samples. This is why, in order to overcome this problem, we had to consider more keys: the right $\tau$, 
large enough to highlight the statistical differences and still  small enough to compute efficiently the corresponding ranks.

\item When we mount our attack we have the freedom to consider for the $\chi^2$ test whatever combination of the ranks $r_j$ we like, as long as random samples are not distinguishable
and AES samples are. We tuned our test to consider only two values (ranks lower than $31743$ and those higher). Two other choices are possible.\\
One would be to look only at even smaller ranks, such as ``ranks lower than $31740$'' (and those higher). We have discarded this option because smaller ranks are very infrequent
and we would then need a very large sample in order to validate our tests.\\
The second choice is to consider more ranks, for example three ranks, as in the following table. 
We considered a total sample having $70$ elements

$$
\begin{tabular}{|c||c c c|}
\hline
\textrm{Rank}&\textrm{Random}& \textrm{AES}& \textrm{Expected}\\
\hline
\hline
        $31745$ &   684191 & 682317 & 683111 \\
     $ 31744$ &  1365480 & 1365113 & 1366222\\
 $\leq 31743$ &    315769 & 318010  &  316107    \\
\hline 
\end{tabular}
$$

According to the $\chi^2$ test we have that:
\begin{enumerate}
 \item Random and Expected, $\rightarrow$ P value equals ${\bf 0.29}$; 
 \item AES and Expected,  $\rightarrow$ P value equals ${\bf 0.0013}$.
\end{enumerate}

So again we would distinguish between random and AES-$128$, but with more difficulty.
This can actually be explained a posteriori: it is true that our embedding would induce less maximum-rank matrices (and the numbers confirm this: $682317 < 683111$), but they
might become $31744$-rank matrices and so add to the most noisy value\footnote{random variable $r_{31744}$ has the largest variance.}, and indeed the $31744$-rank matrices in the AES-$128$ sample are only slighty less than
the expected.
\end{itemize}

We did not report on experiments on the other standard versions of AES (AES-$256$ and AES-$192$), but our preliminary tests seem to indicate  that our test works well also
in those cases, with only a slight worsening of the P value ratio. Indeed, our strategy is independent from the key-length of AES, since our approach is actually independent 
from the key-schedule and only marginally dependent on the round numbers.

\section{Computational effort}
\label{Computational}

The algorithm developed to compute the ranks for the attacks
is specialized for $\FF_2$ and is described in \cite{CGC-cry-prep-enricoAnna10}.
Here we provide a sketch.

Since the matrix is rather large (circa $2^{15}\times 2^{15}$), we must
keep the heaviest part of the computation within the cache.
However, some steps on the whole matrix cannot be avoided, so we need
a strategy that keeps these to a minimum.
In particular, we may need both column and row permutations.
In the below description, we assume that we do not need them.
\begin{remark}
\label{noperm}
To be able to avoid permutations  is equivalent to having performed 
a preprocessing such that each upper-left square block is square and 
non-singular. Of course this cannot be done
a priori, but we stick to this for clarity's sake.
\end{remark}
Without the technicalities pertaing permutations, our algorithm reduces to a variant
of a recursive LU decomposition (\cite{CGC-alg-book-golub96}).
Let the matrix M be in $(\FF_2)^{R\times n}$.
If the whole $M$ does not fit into the cache, we partition M into four
blocks of approximately the same size.

Let $M_1$ be the left-top block. If it does not fit, then we partition
$M_1$ similarly in four blocks and so on.
Let $A$ be the smallest block that does not fit.
We will have
{\small{
\begin{equation}
  \bm{A}
  =
  \begin{pmatrix}
    \bm{A}_{11} & \bm{A}_{12} \\
    \bm{A}_{21} & \bm{A}_{22} \\
  \end{pmatrix} \,,
\end{equation}
}}
where $A_{11}$ fits into the cache.
Thanks to Remark \ref{noperm}, we can assume that block $\bm{A}_{11}$ 
is square and non-singular.
We can  build the $LU$ decomposition of $\bm{A}_{11}=\bm{L}_{11}\bm{U}_{11}$ 
and consider the following equality
{\small{
\begin{equation}
  \bm{A}
  =
  \begin{pmatrix}
    \bm{A}_{11} & \bm{A}_{12} \\
    \bm{A}_{21} & \bm{A}_{22} \\
  \end{pmatrix}
  =
  \begin{pmatrix}
    \bm{L}_{11} & \bm{0} \\
    \bm{A}_{21}\bm{U}_{11}^{-1} & \bm{I} \\
  \end{pmatrix}
  \begin{pmatrix}
    \bm{I} & \bm{0} \\
    \bm{0} & \tilde{\bm{A}}_{22} \\
  \end{pmatrix}
  \begin{pmatrix}
    \bm{U}_{11} & \bm{L}_{11}^{-1}\bm{A}_{12} \\
    \bm{0} & \bm{I} \\
  \end{pmatrix}
\end{equation}
}}
where $\tilde{\bm{A}}_{22} = \bm{A}_{22}-\bm{A}_{21}\bm{U}^{-1}\bm{L}^{-1}\bm{A}_{12}$
is the Schur complement.
Even if block $\tilde{\bm{A}}_{22}$ is singular we can compute the $LU$
decomposition $\tilde{\bm{A}}_{22}=\bm{L}_{22}\bm{U}_{22}$
and the final $LU$ decomposition of matrix $\bm{A}$, as follows:
{\small{
\begin{eqnarray*}
  \bm{A}
  &=&
  \begin{pmatrix}
    \bm{L}_{11} & \bm{0} \\
    \bm{A}_{21}\bm{U}_{11}^{-1} & \bm{I} \\
  \end{pmatrix}
  \begin{pmatrix}
    \bm{I} & \bm{0} \\
    \bm{0} & \tilde{\bm{L}}_{22} \\
  \end{pmatrix}
  \begin{pmatrix}
    \bm{I} & \bm{0} \\
    \bm{0} & \tilde{\bm{U}}_{22} \\
  \end{pmatrix}
  \begin{pmatrix}
    \bm{U}_{11} & \bm{L}_{11}^{-1}\bm{A}_{12} \\
    \bm{0} & \bm{I} \\
  \end{pmatrix}
  =
\\
  &=&
  \underbrace{
  \begin{pmatrix}
    \bm{L}_{11} & \bm{0} \\
    \bm{A}_{21}\bm{U}_{11}^{-1} &\tilde{\bm{L}}_{22}  \\
  \end{pmatrix}
  }_{\bm{L}}
  \quad
  \underbrace{
  \begin{pmatrix}
    \bm{U}_{11} & \bm{L}_{11}^{-1}\bm{A}_{12} \\
    \bm{0} & \tilde{\bm{U}}_{22} \\
  \end{pmatrix}
  }_{\bm{U}} \, .
\end{eqnarray*}
}}
Once we have the LU decomposition of $A$, we use it to recursively compute the LU decomposition of larger blocks containing $A$,
until we reach a global LU decomposition for the whole $M$.
From it, the rank determination is trivial, because it is enough to count the 
nonzeros in the diagonal of $L$.

The three most expensive steps in the above algorithm
are:
\begin{enumerate}
  \item the $LU$ decomposition of block $\bm{A}_{11}$.
  \item the $LU$ decomposition of block $\tilde{\bm{A}}_{22}$.
  \item the computation of the Schur complement $\tilde{\bm{A}}_{22}$.
\end{enumerate}
All three steps cost at most $O(n^3)$, with standard matrix multiplication, however the actual constants differ
and depend on the matrix structure and sparsity.
The cost of matrix multiplication can be lowered theoretically with the Strassen method (\cite{CGC-alg-art-Str69}),
but our matrices are too small to take advantage of it.
However, they are large enough to entice the use of the famous {\it four Russian algorithm} (\cite{CGC-alg-art-4russians}).
We refer to \cite{CGC-cry-prep-enricoAnna10} for our
exact strategy.

\subsection{Cost of the attack}

The attack works very well with $128$ keys, although $70$ are usually enough.
We draw a very conservative estimate of its cost in this subsection.

The attacker needs to collect $2^{16}$ pairs encrypted with the same key.
Since we are requiring $128$ keys, it means that the total number
of pairs is 
$$
      2^{16} \cdot 2^{7} = \; \mathbf{2^{23}} \,.
$$
For any key, the attacker must compute about $2^{15}$ matrix ranks.
Any rank computation costs\footnote{using our algorithm, this is an 
experimental estimate.} $2^{26}$ ecnryptions.
Therefore, the total cost of the attack is
$$
    2^{26}\cdot 2^{15} \cdot 2^{7} = \, \mathbf{2^{48}} \, \mbox{encryptions}
$$

\section{Conclusions}
\label{Conclusioni}
Reduced-round versions of AES-128, AES-192, AES-256 are known to be weak,
although none of these attacks come close to the actual number of rounds.
The best-known attacks use advanced differential cryptanalysis and depend
heavily on the key-scheduling algorithm. Our distinguishing attack is
independent
of the key-schedule and depends only on the round structure. Therefore, it
may be successful even if a huge number of rounds is used.\\
We strongly invite anyone to try our attack, with any number of
rounds,  and we put our software freely usable at 
\begin{center}
 http://www.science.unitn.it/\textasciitilde sala/AES/
\end{center}
The more statistical evidence we collect, the more confidence we will grow
in our results.
Of course, it is possible that a refinement of our approach might lead
to a key-recovery algorithm. Yet, we have not been able to see how, since
the link between the key and the rank statistics is still unclear.

\section*{Acknowledgments}
A preliminary attack with the same underlying philosophy is contained in the first author's Ph.D thesis.
The first author would like to thank the second author (her supervisor).
The computational part contributed by the third author (with very fast implementations) has been
essential to the success of this attack.

We thank G. Naldi and G. Aletti for the use of the computer cluster Ulisse (University of Milan).

These results have been presented in a few talks (2009: Trento; 2010: Marseille, Torino) and several scientific discussions with colleagues. 
The authors would like to thank all the people involved in our discussions, for their valuable comments and suggestions.
In particular, we would like to thank 
T. Mora, L.~Perret and C. Traverso.
%
%
Some discussions took place during the Special Semester on Groebner Bases (2006), organized by RICAM, {\em Austrian Academy of Sciences} and RISC, Linz, Austria.

Part of this research has been funded by: {\tt Provincia Autonoma di Trento grant}{\em ``PAT-CRS grant''}, {\tt MIUR grant}{\em ``Algebra Commutativa, Combinatoria e Computazionale''}, 
{\tt MIUR grant} {\em  ``Rientro dei Cervelli''}.

\bibliography{RefsCGC}

\end{document}